\documentclass[twocolumn,showpacs,preprintnumbers,amsmath,amssymb,prl]{revtex4}

\usepackage{graphicx}
\usepackage{bm}

\begin{document}

\title{Transport coefficients of soft sphere fluids at high densities}

\author{Yu. D. Fomin}
\affiliation{Institute for High Pressure Physics, Russian Academy
of Sciences, Troitsk 142190, Moscow Region, Russia}

\author{V. V. Brazhkin}
\affiliation{Institute for High Pressure Physics, Russian Academy
of Sciences, Troitsk 142190, Moscow Region, Russia}

\author{V. N. Ryzhov}
\affiliation{Institute for High Pressure Physics, Russian Academy
of Sciences, Troitsk 142190, Moscow Region, Russia}

\date{\today}

\begin{abstract}
Molecular dynamics computer simulation has been used to compute
the self-diffusion coefficient,  and shear viscosity  of
soft-sphere fluids, in which the particles interact through the
soft-sphere or inverse power pair potential. The calculations have
been made along the melting line in a wide range of pressures and
temperatures. The validity of scaling relations for thermodynamic
parameters and kinetic coefficients was checked. It was shown that
the Stokes–-Einstein relationship is obeyed if the Barker diameter
is used as a characteristic length scale. It was also shown that
the viscosity is non-monotonic along the isochores as predicted by
Ya. Rosenfeld. It was shown that the viscosity is strongly growing
along the melting line, however, this increase does not stimulate
the glass transition because the relaxation time is decreasing.
\end{abstract}

\pacs{61.20.Gy, 61.20.Ne, 64.60.Kw} \maketitle

\section{I. Introduction}

Knowledge of transport coefficients is extremely important for
characterizing the properties of liquid. They allow one to
characterize how fast the dynamics of the liquid is. Because of
this, transport coefficients are intensively studied
experimentally, theoretically and by simulations. The most
important transport property for an experimental study is
viscosity. Viscosity can easily be measured by a number of
techniques. Nevertheless many questions on the behavior of
viscosity as a function of pressure and temperature still remain.
Some of them are considered in the present article.

It is well known that the viscosity of most liquids grows
exponentially with increasing pressure at constant temperature and
demonstrates critical behavior in the vicinity of the glass
transition line. In the case of hard or soft spheres the glass
transition at high pressures corresponds to the jamming of the
system. However, such kind of behavior is typical for supercooled
liquids while the behavior of viscosity close to the melting line
is not clear.

As was discussed in \cite{brazhkin2008}, there are two empirical
approaches to the description of the viscosity behavior of simple
liquids under simultaneous changes of temperature and pressure.
The first approach goes back to Bridgman's works \cite{1} and
suggests that the viscosity of the liquid is nearly constant along
the isochores. The other and currently more widespread approach
was formulated by Pourier \cite{2}. It is based on the presumption
that the viscosity of the melts is invariable along the melting
curve. The experimental data analysis shows that both approaches
are incorrect; the constant viscosity lines for rare gas liquids,
as well as for liquid metals, have a slope that is intermediate
between the melting line and the isochore slopes \cite{brlyap}
(figure 5). As a consequence, the viscosity of simple melts along
the melting curve increases and this increase can be extrapolated
to megabar pressures \cite{brlyap}. However, this extrapolation is
in fact difficult because of a rather small viscosity variation in
the covered range \cite{4}. For example, the viscosity of the Fe
melt grows along the melting curve by several times with the
pressure increase to $10 - 15$ GPa \cite{4,5}. However, the
obtained data do not permit drawing any conclusions with regard to
the character of this growth; various analytic models equally
closely descriptive of the studied baric dependencies give
different extrapolation results. In consequence, different
equations employed in the extrapolation of the Fe melt viscosity
to pressures of $1.4 - 3.1$ Mbar, corresponding to the conditions
existing in the Earth's outer core, provide the data that vary
over  $10 - 20$ orders of magnitude \cite{4} (figure 6). The same
is valid for other 'simple' melts like liquid Ar.

Thus, it is presently impossible to draw inferences about how high
the viscosity can grow along the melting curve for the given melts
with the pressure increase to about $100 - 1000$ GPa. If the
viscosity growth is minor, liquid metals and rare gas liquids will
remain very poor glass formers at megabar pressures. On the other
hand, the viscosity growth along the melting curve amounts to
several orders of magnitude, so the question arises whether the
liquid metals and rare gas liquids become viscous glass-forming
systems under the increasing pressure. In this case, the glass
transition of the given melts in the megabar range should be
similar to the jamming of soft spheres.

The main purpose of this paper is the computer simulation study of
the behavior of the transport coefficients of a soft sphere system
in a wide range of densities and temperatures. Although the
interatomic potential in liquid metals can not be described by a
pair potential, the potential in rare gas liquids can be
approximated by the Lennard-Jones function which can be
substituted in the high density and temperature limit  by a soft
sphere potential. In this respect it is interesting to carry out a
systematic study of the transport coefficients of soft sphere
system as a generic model for real liquids at high pressure.

Note that most of previous studies of transport coefficients of
liquids were concerned to the low density - low temperature region
while high density - high temperature properties were beyond the
consideration. Only few works on the high density transport
properties are available (see, for example, \cite{bastea}) In this
sense the present work is directed to fill this gap.

\section{II. Soft Spheres: Theoretical Approach}

Interestingly, one can make some important predictions about
thermodynamic properties of soft spheres on the basis of the form
of the potential only. To do this one can use the Klein theorem
\cite{klein,klein1}.

Consider the dimensionless density $\rho \frac{\sigma^3}{V}$ and
reduced coordinates $s=r (\frac{N}{V})^{1/3}$. The partition
function of soft spheres in these variables has the following form
\cite{stishov}:

\begin{equation}
  Z=\frac{V^N}{N \! \lambda ^{3N}} \int_{s_1}... \int_{s_N} exp[-\beta
  \rho ^{n/3} \sum_{i<j} s_{ij}^{-n} ] d s_1 ...d s_N,
\end{equation}
where $\lambda=\left(\frac{\hbar}{2 \pi mk_BT}\right)^{1/2}$ is de
Broile wave length and $\beta=1/(k_BT)$. The Klein theorem states
that if $U$ is a homogeneous function of degree $n$, i.e.
$U(\lambda r_1,...,\lambda r_N)=\lambda ^{n} U(r_1,...,r_N)$ then
for $N$ sufficiently large \cite{klein1}

\begin{equation}
  \int_{r_1}...\int_{r_N} exp[-\beta \cdot U(r_1,...,r_N)] dr_1...dr_N=G(\beta \cdot \rho
  ^{-n/3}),
\end{equation}
where $G$ is some function of $(\beta \rho ^{-n  /3})$.

Using the Klein theorem one can find that the equation of state
has the form:

\begin{equation}
  \frac{PV}{Nk_BT}=1+ \phi (\rho
  (\frac{\varepsilon}{k_BT})^{3/n}).
\end{equation}

If the system with such equation of state has a first order phase
transition (let us call it melting), then the transition line is
characterized by the universal parameter
$\rho(\frac{\varepsilon}{k_BT})^{3/n}$ \cite{stishov}. From this
it follows that the densities of a liquid and crystal along the
melting line are:

\begin{equation}
  \rho_l=c_l (\frac{k_BT}{\varepsilon})^{3/n}
\end{equation}

\begin{equation}
  \rho_c=c_c (\frac{k_BT}{\varepsilon})^{3/n},
\end{equation}
where $c_l$ and $c_c$ are constants. One can see that the
densities along the melting line follow some scaling law.

This approach was further developed in the work \cite{jzah}, where
the scaling relations were derived under the assumption of scaling
invariancy of the Hamilton equations for homogeneous potentials.

Remember that kinetic energy is a homogeneous function of the
power $2$:

\begin{equation}
 K(ap_i)=a^2 K(p_i)
\end{equation}

If scaling $r_i=ar_i$, $p_i=bp_i$ and $t=ct$ with $b^2=a^{-n}$ and
$c^2=a^{2+n}$ is applied then the new phase trajectory of the
system has the same geometric shape \cite{jzah}. Then the
following scaling laws are valid (we skip the Boltzmann constant
in the formulas below for the sake of brevity):

\begin{equation}
  V=a^3V_0
\end{equation}

\begin{equation}
  T=\frac{2}{3N}\langle \sum \frac{p_i^2}{2m_i} \rangle
  \longrightarrow T=a^{-n} T_0 \longrightarrow V=V_0 \left(\frac{T}{T_0}\right)^{-3/n}
\end{equation}

\begin{equation}
  P=\frac{NT}{V}-\frac{1}{3V}\langle \sum q_i \frac{\partial U}{\partial q_i}
  \rangle \longrightarrow P=P_0 \left(\frac{T}{T_0}\right)^{1+3/n}.
\end{equation}

Using the Green - Kubo relations for the diffusion
$D=\int_{0}^\infty <\textbf{V}(t)\textbf{V}(0)>dt$ and viscosity
$\eta=\frac{1}{Vk_BT} \int_0^\infty <\sigma^{xy}(t)
\sigma^{xy}(0)>$, where $\sigma^{xy}$ is a stress tensor
component, one can derive the scaling relations for diffusion and
viscosity:

\begin{equation}
\eta \sim P^{\frac{n+4}{2n+6}}
\end{equation}
or
\begin{equation}
\eta \sim T^{\frac{n+4}{2n}},
\end{equation}
and
\begin{equation}
D \sim P^{\frac{n-2}{2(n+3)}}
\end{equation}
or
\begin{equation}
D \sim T^{1/2-1/n}.
\end{equation}

Below we compute different quantities in simulation and fit the
results to the corresponding scaling law using the least square
procedure.

\begin{figure}
\includegraphics[width=7cm]{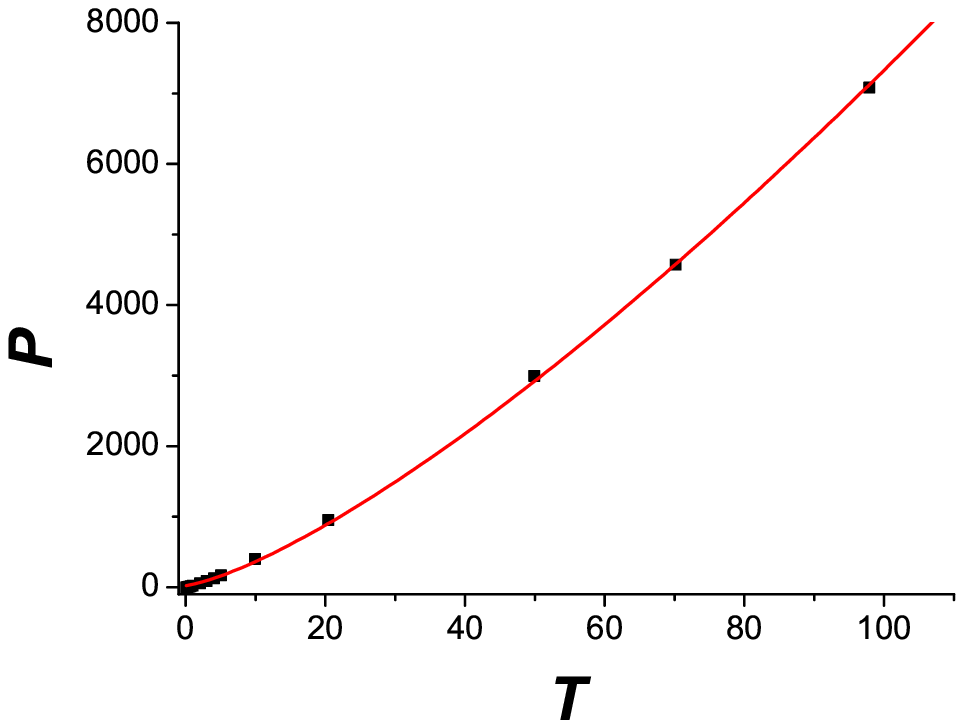}%

\includegraphics[width=7cm]{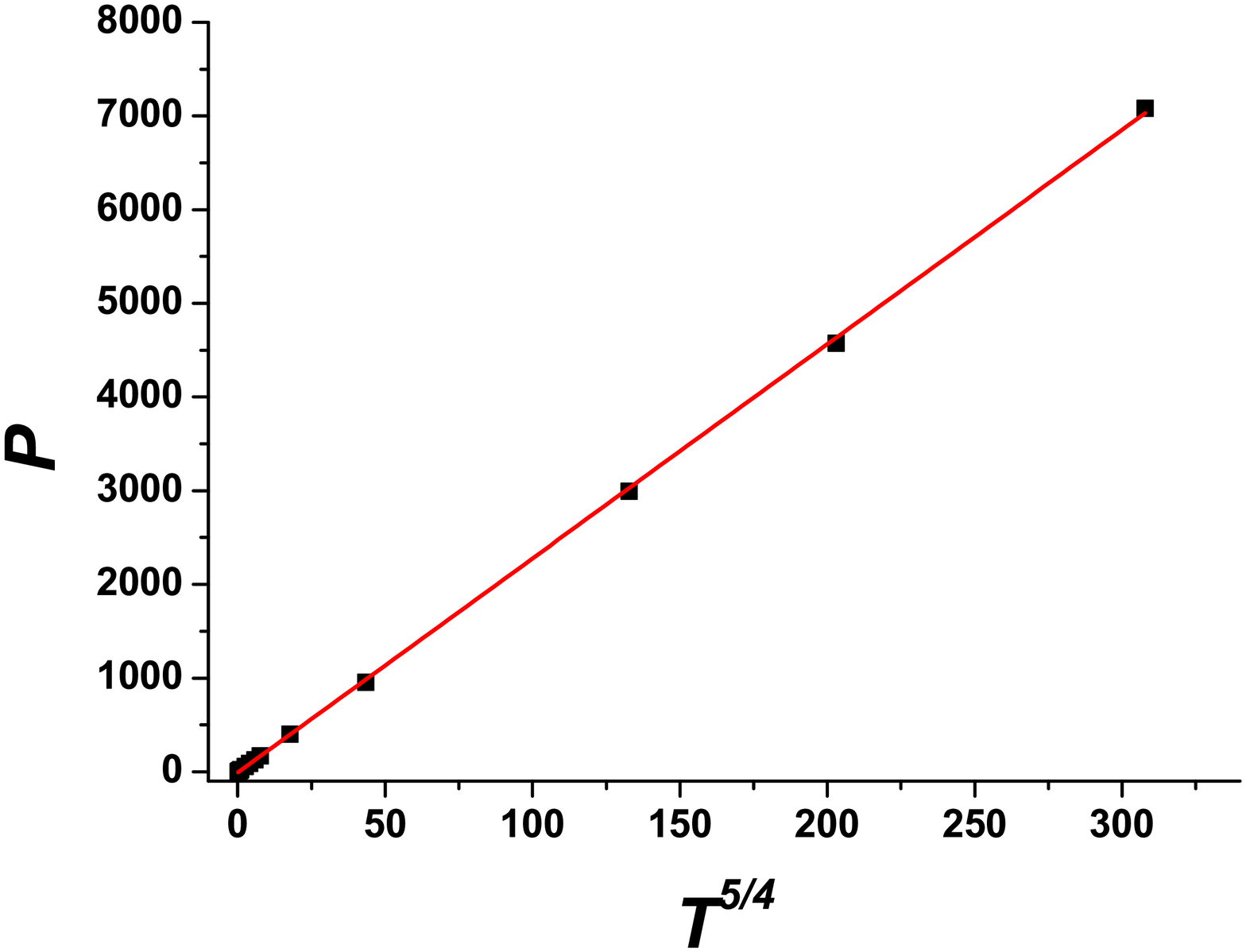}%

\caption{\label{fig:fig1b} Pressure as a function of temperature
along the melting line in usual (a) and scaled (b) forms. The
symbols are MD data and the line is the scaling law fitting.}
\end{figure}

One can see that, following \cite{jzah}, the scaling relations can
be generalized to be applicable to the transport coefficients,
however this approach involves some assumptions which are not
fully proved. This is why it is particularly interesting to check
whether the scaling predictions for the transport coefficients are
correct.

\section{III. Simulation}

In the present work we study the viscosity, $\eta_s$,  and
diffusivity, $D$, of soft spheres. The choice of the system is
determined by the following reasons. In order to investigate the
behavior of the system along the melting line in a wide range of
pressures we need to know the phase diagram of the system
including a high pressure region. In most cases the melting lines
were studied in the vicinity of the triple points.  As it was
shown that the soft sphere system obeys the scaling law. The
parameters of the scaling of pressure and density with temperature
along the melting line were determined both from simulation
\cite{hrj} and from density functional theory \cite{dftsoft}.

The soft sphere potential has the form:
\begin{equation}
  \Phi(r)=\varepsilon \left(\frac{\sigma}{r}\right)^{n}.
\end{equation}

In our simulations we use $n=12$. The potential is cut at the
distance $r_c=2.5$ as in standard simulations of Lennard-Jonse
systems. Reduced units ($\varepsilon=1$ and $\sigma=1$) are used
in the paper. As it was shown (see, for example, previous section
and \cite{hrj}), the behavior of soft spheres system depends on
the parameter $\gamma_n=\rho \sigma^3
(\frac{k_BT}{\varepsilon})^{-3/n}$. In \cite{hrj} it was shown
that the line of fluid to solid transition is $\gamma_n=1.15$.
From this formula one can compute the transition density to be
used in simulations.

The system of 1000 particles was simulated in the microcanonical
(constant N, V and E) ensemble. Equations of motion were
integrated by the velocity Verlet algorithm. The equilibration
time was set to $5\cdot 10^5$ time steps and the production time -
$3.5\cdot 10^6$ steps. The time step we use is $dt=0.0005$. During
equilibration the velocity rescaling was applied to keep the
temperature constant. During the production cycle, the $NVE$
ensemble was used. The diffusion coefficient was calculated from
the mean square displacement via Einstein relation, while the
viscosity was determined by the integration of the stress
correlation function \cite{bib}.

The computation of the phase diagram of the system was also done.
The reason for this computation is as follows. To our knowledge,
all previous works studying soft sphere phase diagram considered
the system at temperatures of the order of unity. Although it
allows finding the scaling coefficient $\gamma_n$, the value may
contain some errors. Since in the present work we study the
temperatures up to $T=100$, these small errors can give quite
large uncertainty in the location of the melting line.

Because of this we computed the melting densities at $T=20$ and
$T=50$. The same system of 1000 particles was used for this
calculation in the case of the liquid and of $1372$ particles for
the FCC solid. The transition points were determined via the
double tangent construction to the free energy lines. The crystal
free energy was computed by coupling to the Einstein crystal
\cite{bib}, while the liquid free energy was determined by
integrating the equation of state from the dilute gas limit
\cite{bib}. Table 1 presents a comparison of the melting points
obtained by our simulations to the predictions based on the
scaling factor from Ref. \cite{hrj}.

\begin{tabular}{|c|c|c|c|}
\hline
$T$&$\rho_l$ (this work)&$\rho_l$ (ref. \cite{hrj})&relative discrepancy\\
\hline
20&2.45&2.43&$0.8\%$\\
50&3.12&3.06&$1.6\%$\\
\hline
\end{tabular}

We can see that the agreement is rather good. It confirms that one
can use the scaling parameters from the reference \cite{hrj}.

\section{IV. Results and Discussion}

\subsection{Transport Coefficients Behavior}

The main purpose of this paper is to study the behavior of
transport coefficients - shear viscosity and diffusion  - in a
wide range of densities and temperatures. To do this, we measure
transport coefficients along the melting line and along a set of
isochores for temperatures ranging from $T=0.1$ up to $T=100.0$.
The following isochores are considered: $\rho =0.65; 0.95; 1.51;
2.43; 2.69; 2.89; 3.06; 3.2; 3.33; 3.44$. Note that the density
and temperature dependence of pressure is rather strong; thus, the
investigated density and temperature range corresponds to the
pressure range varying from about unity to about $7000$. Changing
Lennard-Jones reduced units to real ones, we have the pressure
varying in the range up to $300$GPa. It covers almost the entire
range of experimentally attainable static high pressures.

\begin{figure}
\includegraphics[width=7cm]{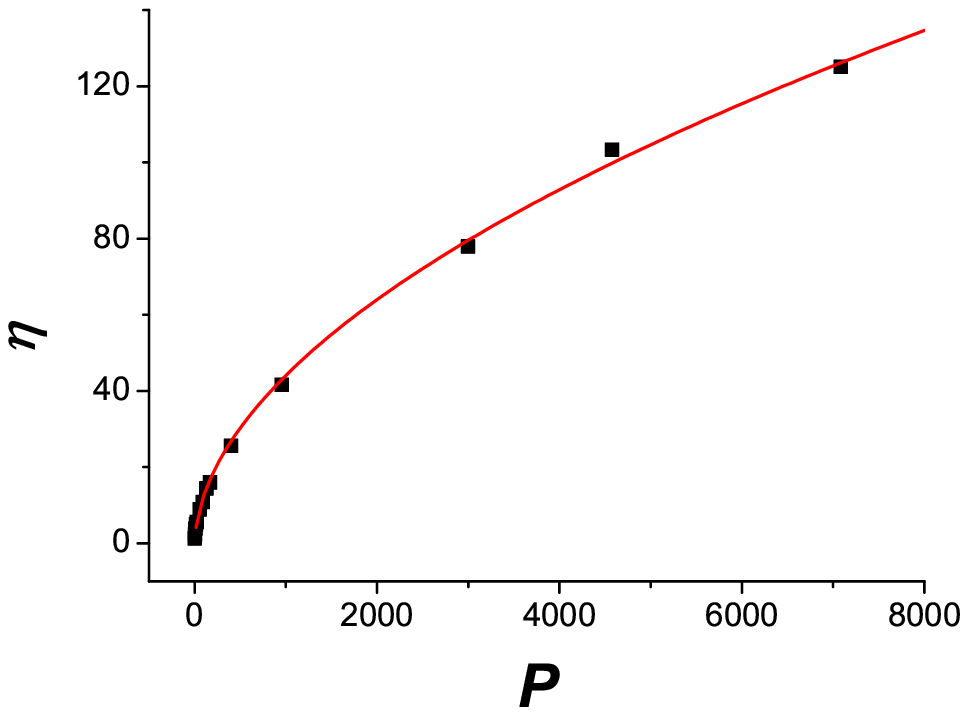}%

\includegraphics[width=7cm]{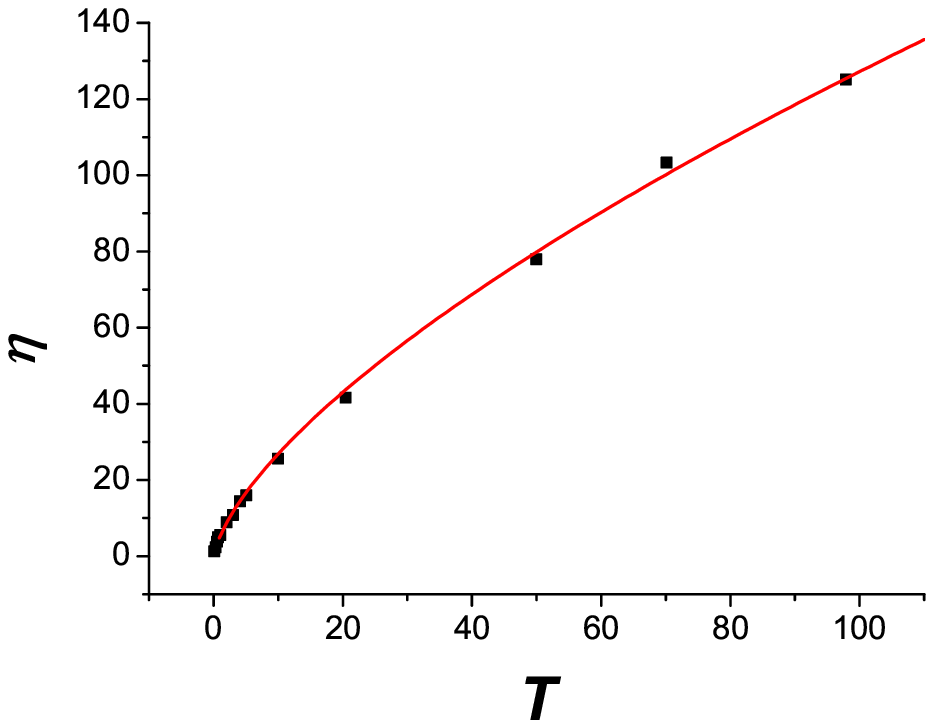}%
\caption{\label{fig:fig2b} Viscosity along the melting line as a
function of pressure and temperature. The symbols are the MD data
and the line is the scaling law fitting.}
\end{figure}

\begin{figure}
\includegraphics[width=7cm]{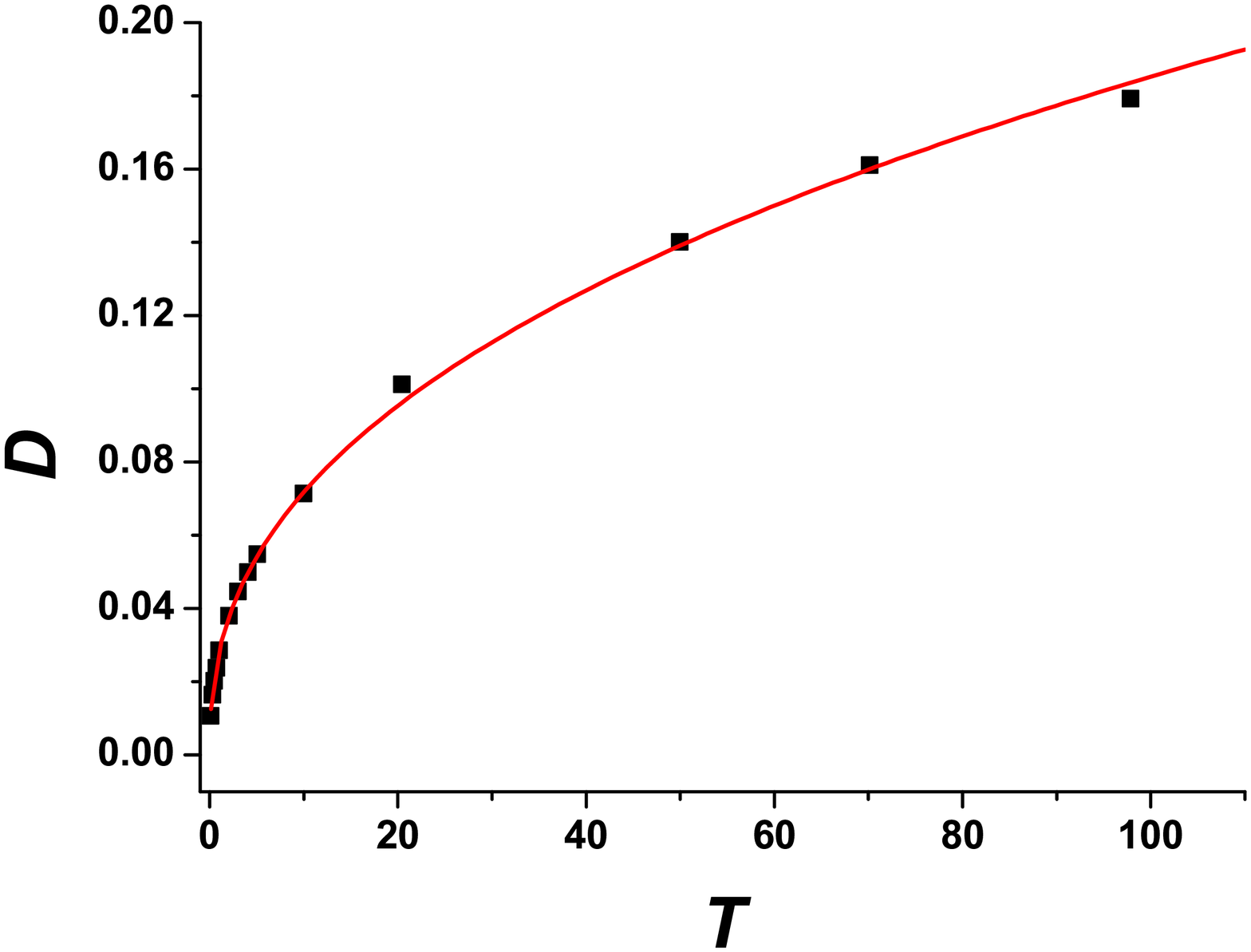}%

\includegraphics[width=7cm]{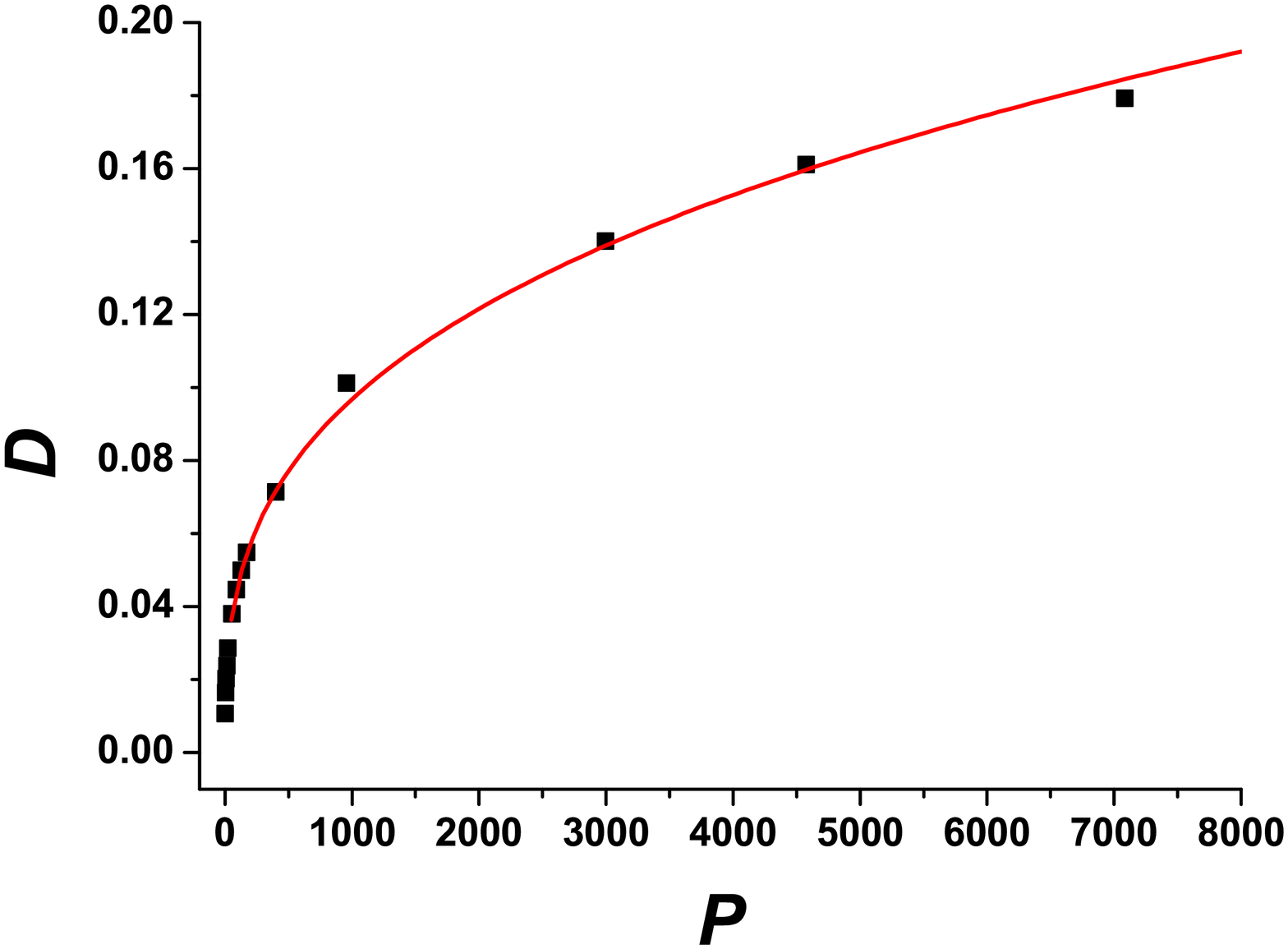}%
\caption{\label{fig:fig3b} Diffusion coefficient along the melting
line as a function of pressure and temperature . The symbols are
the MD data and the line is the scaling law fitting.}
\end{figure}

Figs. 1a and 1b present the pressure dependence along the melting
line in usual and scaled forms. One can see that the scaling law
works well.

The next figures (Figs. 2a and 2b) show the behavior of viscosity
along the melting line as functions of both pressure and
temperature and the fitting to the corresponding scaling formulas.
As one can see, the scaling rules for viscosity are well fulfilled
too. Unlike some predictions, the viscosity is not constant along
the melting line. The mechanism of the viscosity growth will be
discussed below.

Finally, Figs. 3a and 3b present the diffusion coefficient along
the melting line.

Figs. 2 and 3 show the diffusion and viscosity growth along the
melting line. Although it may seem counter intuitive, they satisfy
to the scaling laws. We will discuss this problem below.

Figs.4a and 4b show the diffusion coefficient along different
isochores and along the melting line as functions of temperature
and pressure. As one can expect the diffusion increases upon
temperature growth while at fixed temperature it decreases with
increasing density. Diffusion also increases with increasing
pressure along the isochores because in this situation we have
simultaneous increase of temperature.

The behavior of viscosity along the isochores is more complex
(Figs. 5a and 5b). One can see from the insets in the figures that
the viscosity is not a monotonic function of pressure and
temperature. The viscosity decreases with pressure close to the
melting line, and increases away from it.  Another conclusion from
these figures is that the viscosity behaves qualitatively similar
along different isochores - the shapes of the curves are almost
the same for different densities.

A possible explanation of the nonmonotonic behavior of the
viscosity was suggested by Rosenfeld \cite{rosenfeld}. In his
works \cite{rosenfeld,ros1} Rosenfeld showed the connection
between transport coefficients of a fluid and excess entropy.
Considering the behavior of the entropy he showed that the
viscosity can have a minimum. He also gave a qualitative
explanation of this fact. If we consider viscosity as an integral
over stress correlation function, we can split it to three
contributions: kinetic - kinetic ($kk$), kinetic - potential
($kp$) and potential - potential ($pp$)\cite{bib,hansmcd}:

\begin{equation}
\eta=\int_0^{\infty} C_{kk}(t)dt+\int_0^{\infty}
C_{kp}(t)dt+\int_0^{\infty} C_{pp}(t)dt,
\end{equation}
where
\begin{equation}
 C_{kk}(t)=\frac{m^2 \rho}{Nk_BT}\sum_{i=1}^N \sum_{j=1}^N
 <v_{x,i}(0)v_{y,i}(0)v_{x,j}(y)v_{y,j}(y)>,
\end{equation}
where $v_{x,i}(t)$ denotes x component of the velocity of the
particle $i$ at time $t$.
\begin{widetext}
\begin{equation}
 C_{pk}(t)=\frac{2m \rho}{Nk_BT}\sum_{i=1}^N \sum_{j=1}^N \sum_{k=1, k\neq
 j}^N\langle v_{x,i}(0)v_{y,i}(0)\cdot
 \frac{x_{jk}(t)y_{jk}(t)}{r_{jk}(t)^2}F[r_{jk}(t)]
 \rangle,
\end{equation}
\end{widetext}
where $x_{jk}(t)$ and $y_{jk}(t)$ are Cartesian components of
$\textbf{r}_{jk}(t)$ and $F(r)=-\frac{r}{2} \frac{\partial
\Phi[r]}{\partial r}$ - force between particles $j$ and $k$.

Finally
\begin{widetext}
\begin{equation}
 C_{pp}(t)=\frac{\rho}{Nk_BT}\sum_{i=1}^N \sum_{j=1, j\neq i}^N
 \sum_{k=1} \sum_{l=1, l\neq k}^N \langle \frac{x_{ij}(0)y_{ij}(0)}{r_{ij}(0)^2}F[r_{ij}(0)]\cdot
\frac{x_{kl}(t)y_{kl}(t)}{r_{kl}(t)^2}F[r_{kl}(t)].
 \rangle.
\end{equation}
\end{widetext}

The role of different contributions to the viscosity for a Lennard
- Jones fluid was studied by several authors (see, for example,
\cite{pc1, pc2}).

\begin{figure}
\includegraphics[width=9cm]{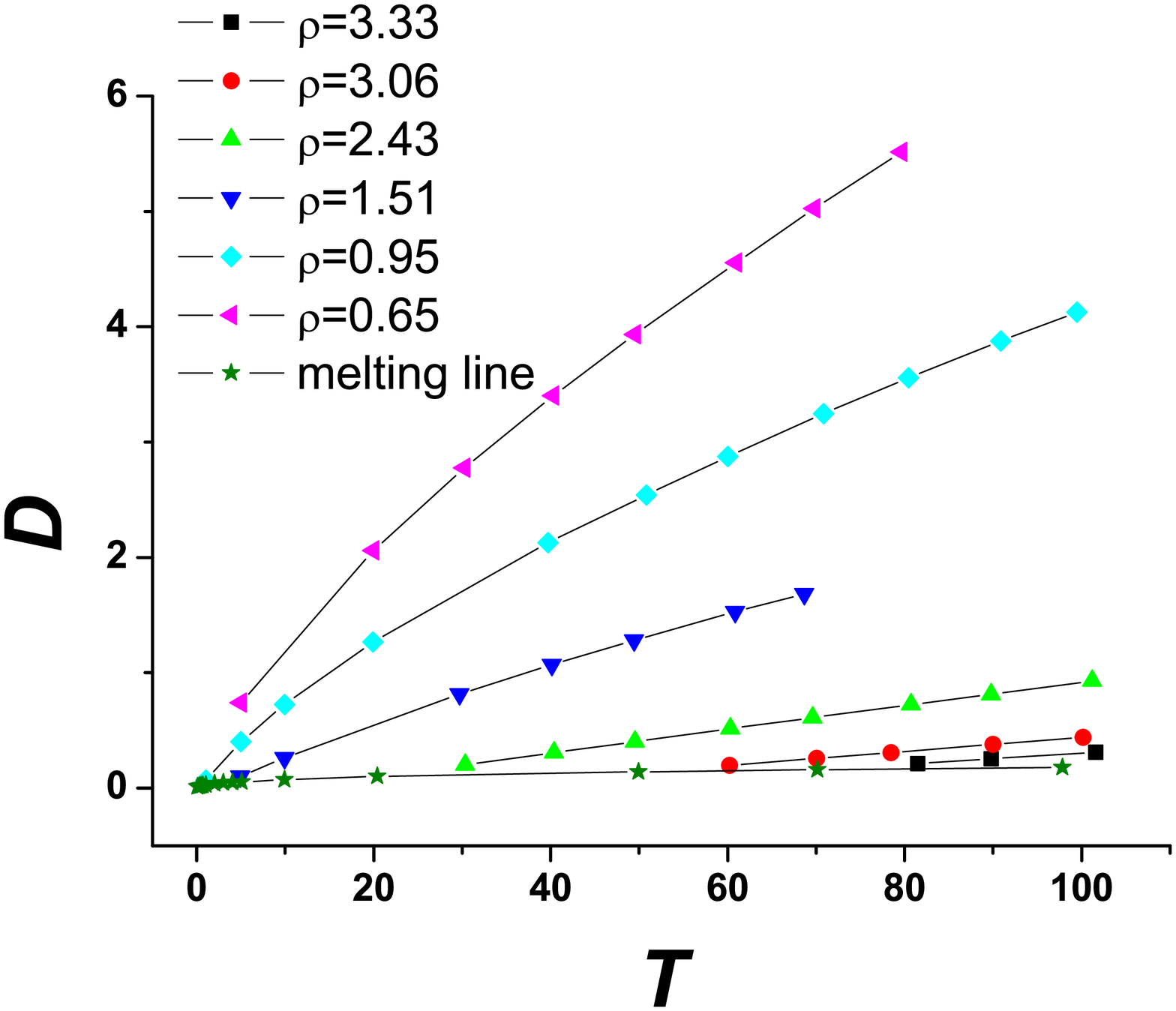}%

\includegraphics[width=9cm]{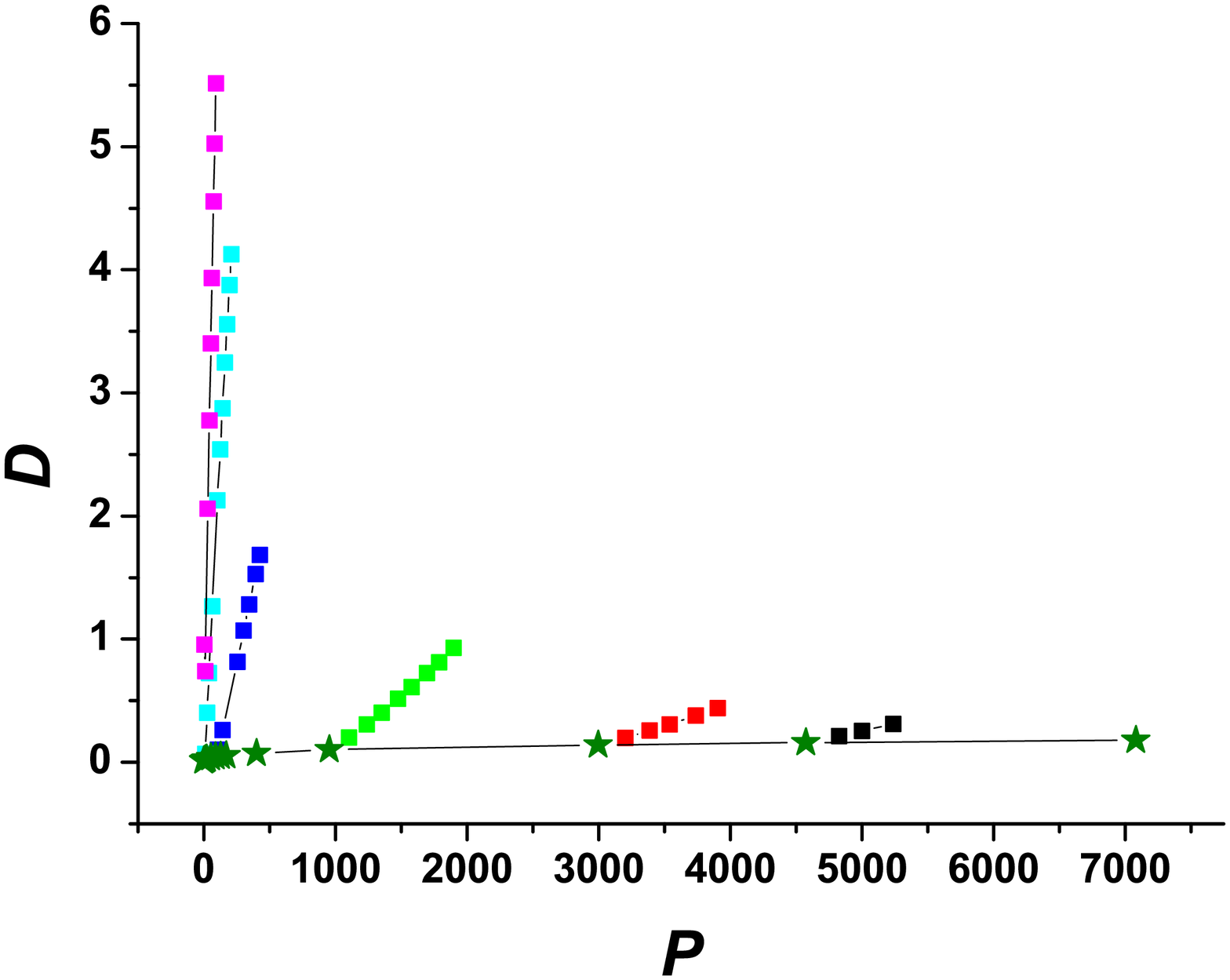}%
\caption{\label{fig:fig4} Diffusion coefficient along several
isochores and the melting line as functions of temperature and
pressure. (Color online)}
\end{figure}

\begin{figure}
\includegraphics[width=9cm]{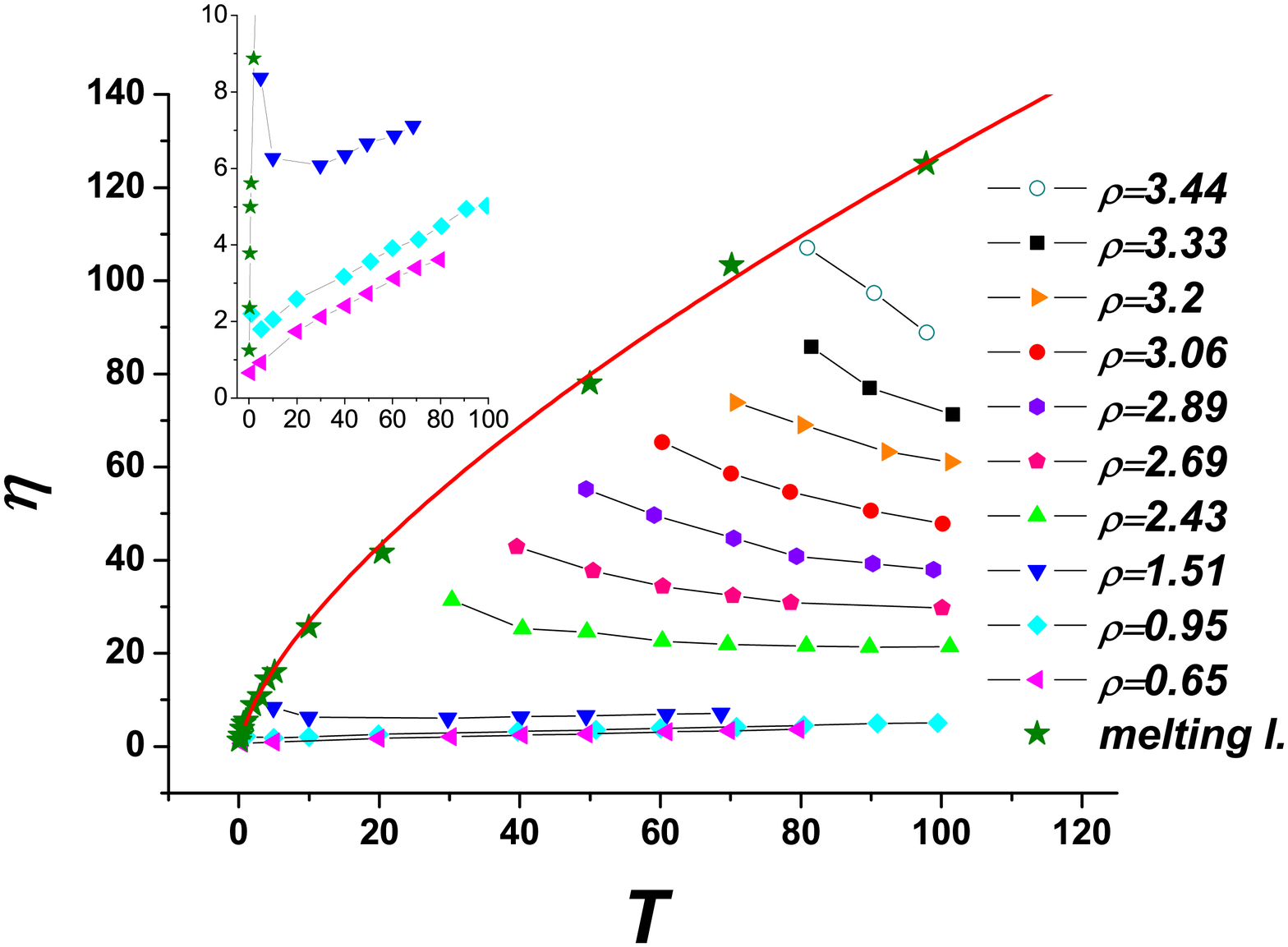}%

\includegraphics[width=9cm]{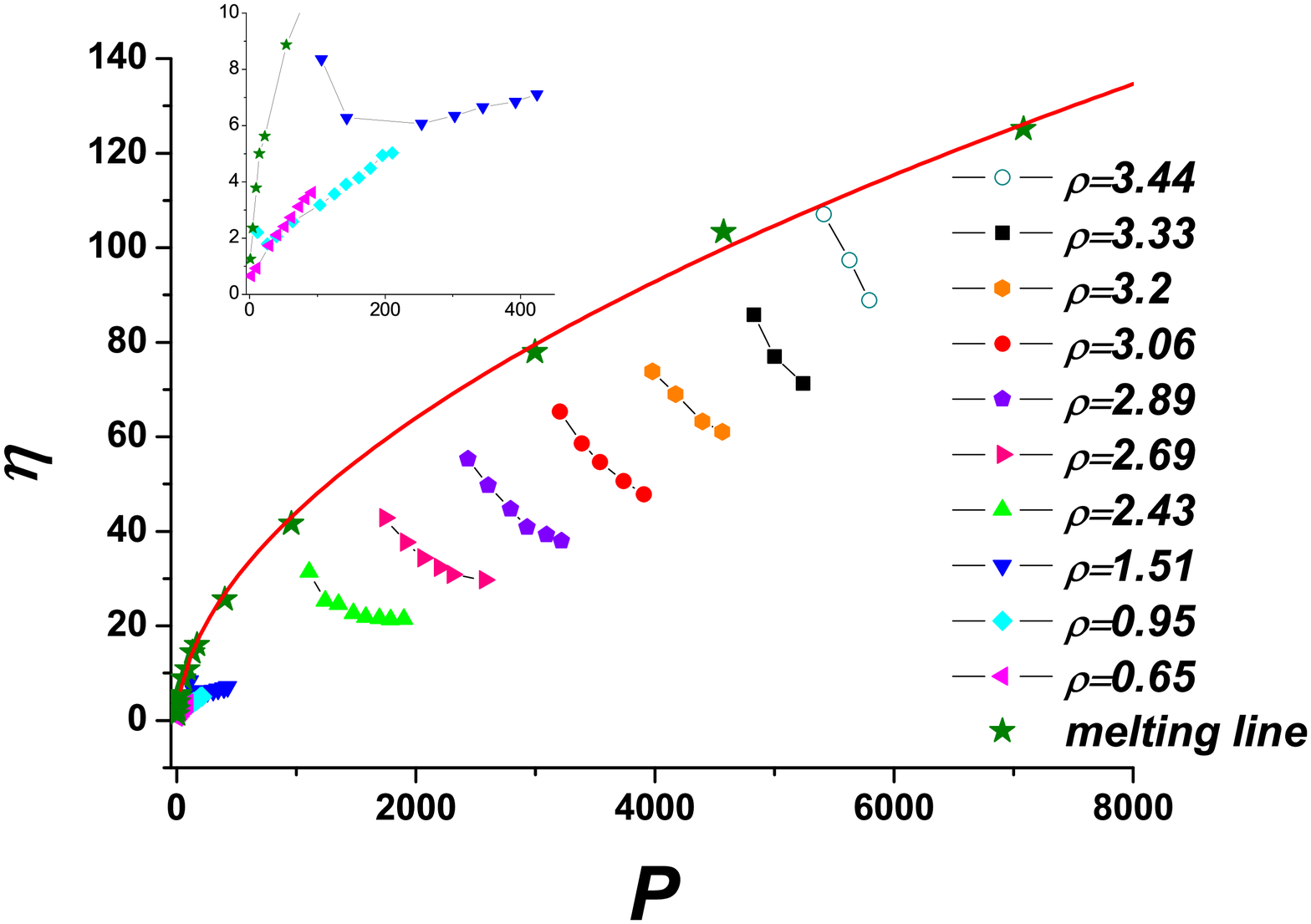}%

\caption{\label{fig:fig5b} Viscosity along the isochores (squares)
and along the melting line (stars)as a function of pressure and
temperature. The symbols are MD data and the line is the scaling
law fitting. The insets represent the low pressure or low
temperature behavior of the plots.(Color online)}
\end{figure}

One can expect that close to the melting line the main
contribution to the viscosity comes from the potential - potential
correlation, while at high temperatures the kinetic - kinetic part
should be dominant. Having the $pp$ correlation decreasing and the
$kk$ one increasing, we can expect that a minimum of the total
viscosity can be observed.

\begin{figure}
\includegraphics[width=10cm]{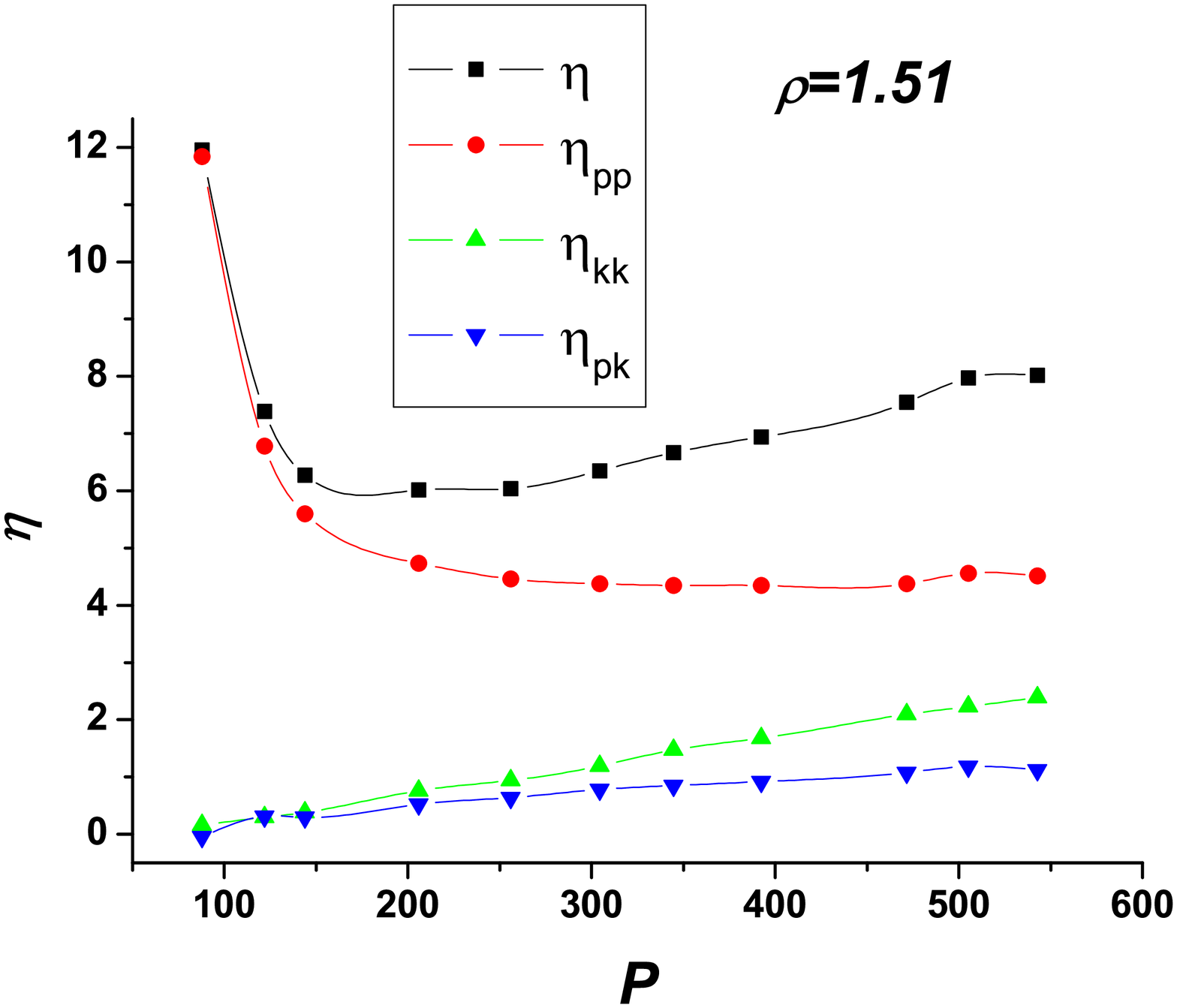}%

\includegraphics[width=10cm]{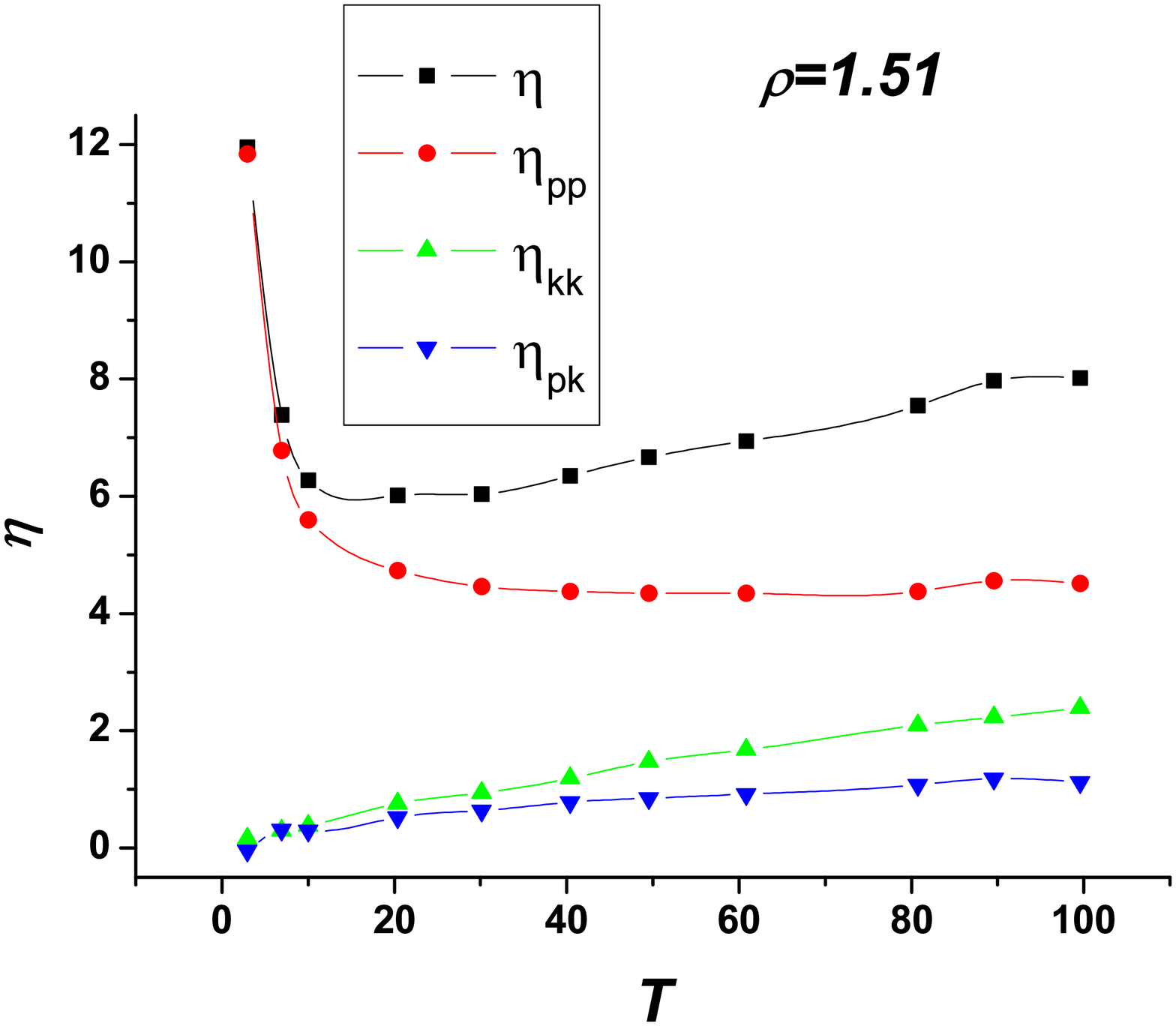}%
\caption{\label{fig:fig6b} Different contributions to viscosity as
a function of pressure and temperature along the isochore
$\rho=1.51$. Squares - full viscosity, circles - $pp$
contribution, up triangles - $kk$ contribution and down triangles
- $kp$ contribution.(Color online)}
\end{figure}

In order to verify this suggestion we carried out the calculations
of the different contributions to the viscosity for the density
$\rho=1.51$ and a set of temperatures (pressures) (Figs. 6a and
6b). As one can see from these figures the suggestion is correct:
while the $pp$ contribution decreases with increasing temperature
both the $kk$ and $kp$ ones increase. Because of this one can
expect the further increase of the viscosity with increasing
temperature.

Given this increase of the viscosity, one can pose a question
about a glass transition in the soft sphere system. Two
definitions of the glass transition point are common: the one
which is connected to the relaxation time of the system, and
another which defines the glassification through the viscosity
increase. The Maxwell relaxation time and shear viscosity are
related via the equation \cite{hansmcd}:
\begin{equation}
  \tau=\eta /G_{inf}
\end{equation}
where
\begin{equation}
  G_{inf}=\rho k_BT+ \frac{2 \pi \rho ^2}{15}\int_0 ^{\infty} dr
  g(r) \frac{d}{dr} (r^4 \frac{d\Phi}{dr})
\end{equation}
is the infinite-frequency shear modulus \cite{ginf1,ginf2}.

\begin{figure}
\includegraphics[width=8cm, height=8cm]{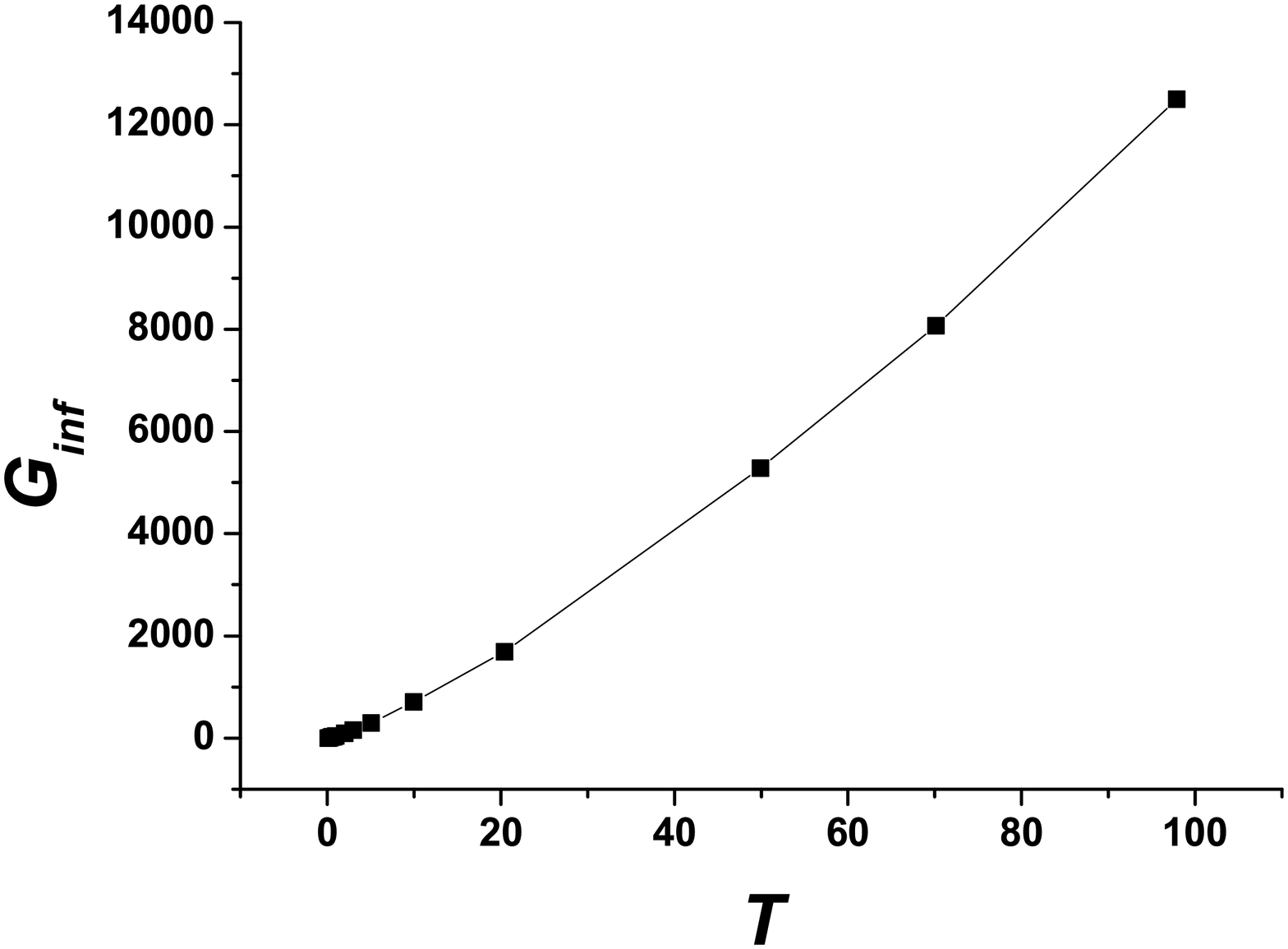}%
\caption{\label{fig:fig6b} Infinite-frequency shear modulus
$G_{inf}$ along the melting line.}
\end{figure}

\begin{figure}
\includegraphics[width=8cm, height=8cm]{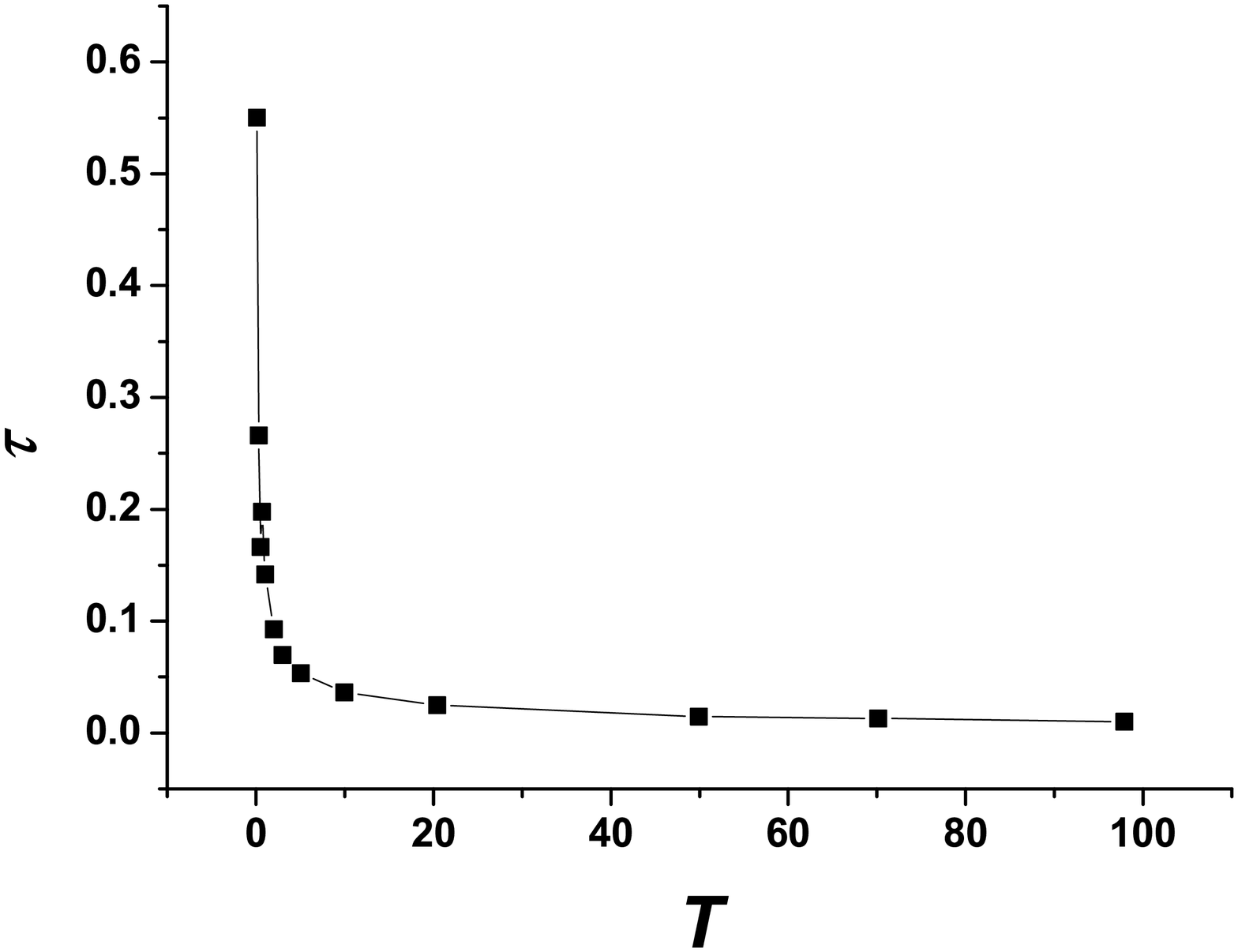}%
\caption{\label{fig:fig6b} Relaxation time along the melting
line.}
\end{figure}

The behavior of the elastic modulus $G_{inf}$ along the melting
line is shown in Fig.7. One can see that $G_{inf}$ grows
dramatically.  Fig. 8 presents the relaxation time along the
melting line. As it is clear from this plot, although the
viscosity of the system increases with increasing temperature
along the melting line the relaxation time becomes very small.
Having small relaxation time, the system can definitely not become
glasslike.

One can conclude that the main glassification criterion is the one
through the relaxation time. As a result, even at certain
conditions one has the viscosity increase, at the same time the
high viscosity does not mean glassification because of decrease of
the relaxation time. It seems that it is a high $pp$ viscosity
that brings the system to glass transition while dominating $kk$
viscosity does not.

\subsection{Stokes - Einstein Relation}

Given the behavior of the viscosity and diffusion coefficient, it
is interesting to consider the Stokes - Einstein relation. It can
be written as
\begin{equation}
 c=\frac{k_BT}{\pi\sigma D \eta},
\end{equation}
where $\sigma$ is the diameter of a particle and $c$ is the
constant which equal 2 for "stick" boundary conditions and 3 for
the "slip" ones \cite{hansmcd}. Several authors studied the Stokes
- Einstein coefficient $c$ in different systems \cite{zwanzig,
seliqmet, heyes, heyes1, heyes2}. The deviation from the Stokes -
Einstein law was found. In Reference \cite{zwanzig}, a modified
relation was suggested. To take into account the effect of a
change of the diameter $\sigma$, it was suggested to use the
relation:

\begin{equation}
 c=\frac{k_BT}{\pi\sigma D \eta \rho^{1/3}}. \label{c}
\end{equation}

Although this normalization reduced the deviation, it did not make
the coefficient $c$ constant. The study of the Stokes - Einstein
relation in the soft sphere system \cite{heyes, heyes1, heyes2}
also confirmed the deviation from both above formulas.

Here we apply the scaling laws to understand the Stokes - Einstein
coefficient behavior. Recall that along the melting line the
diffusion coefficient changes as $D\sim T^{5/12}$ while $\eta \sim
T^{2/3}$. Inserting it in Eq. (\ref{c}) we obtain $c \sim
T^{-1/12}$. Figs. 9a and 9b represent the Stokes - Einstein
coefficient $c$ along the melting line in usual and scaled forms.

\begin{figure}
\includegraphics[width=7cm]{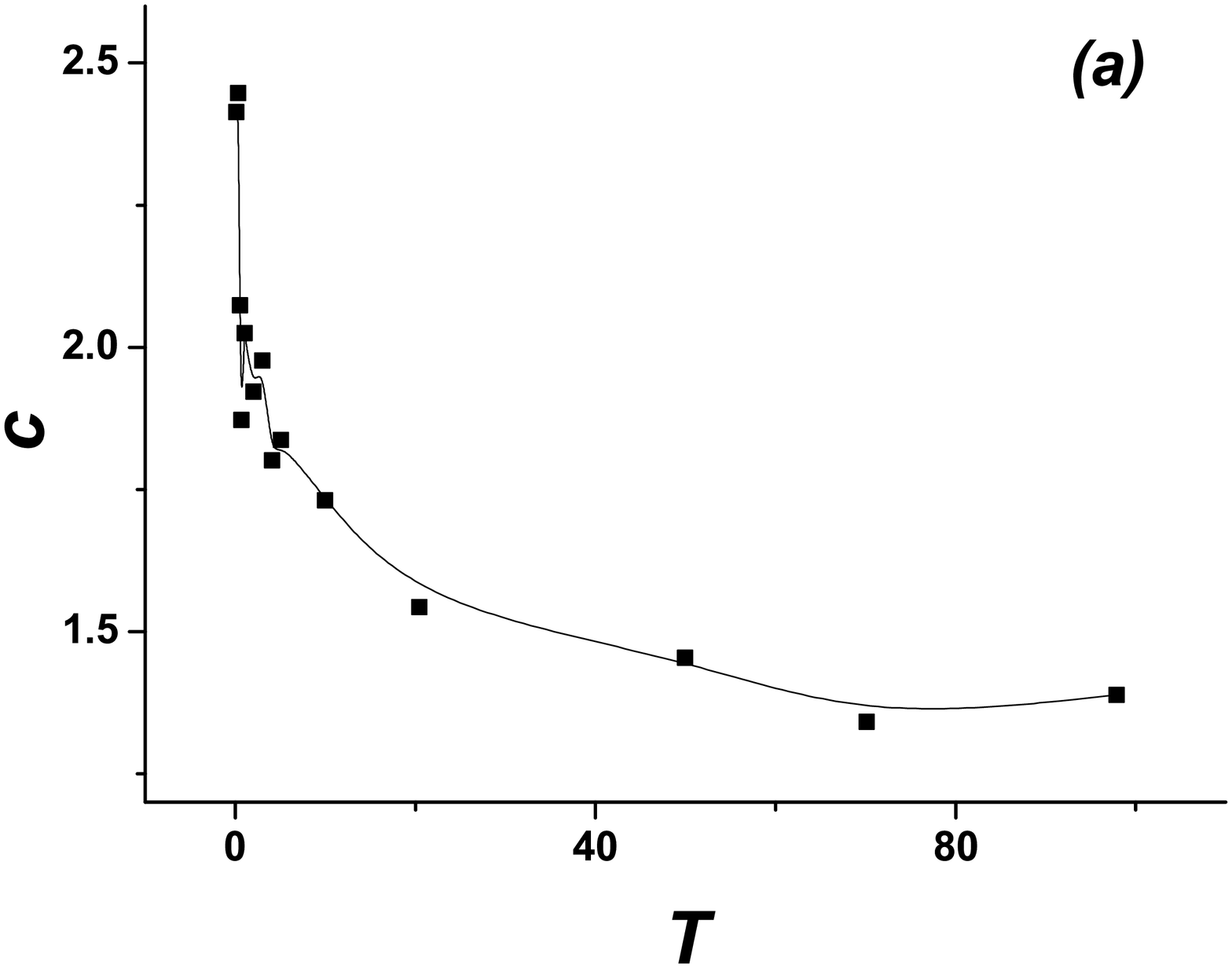}%

\includegraphics[width=7cm]{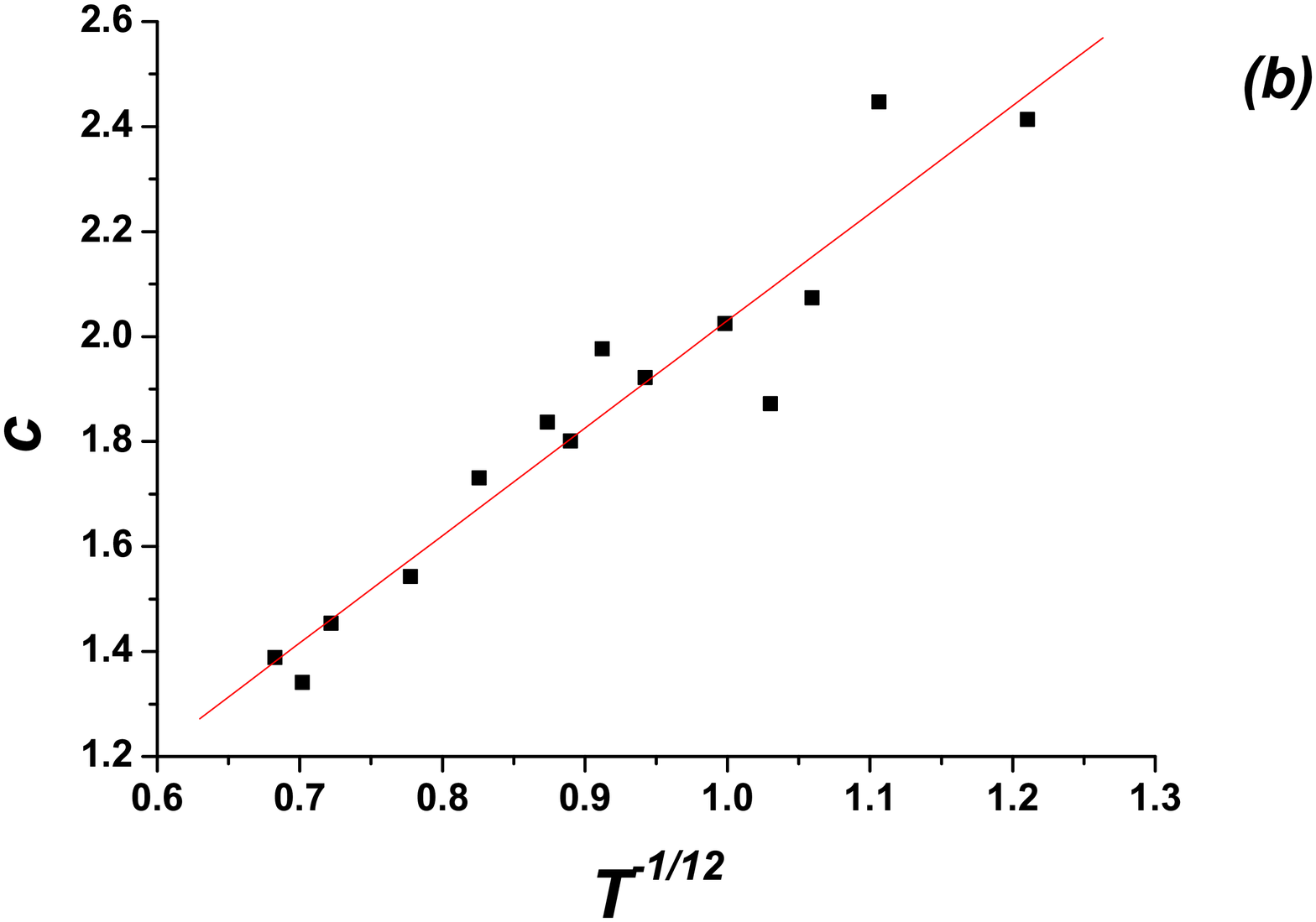}%
\caption{\label{fig:fig7b} Stokes - Einstein coefficient $c$ in
usual (a) and scaled (b) forms. Squares - MD results, line (b) is
linear fit.}
\end{figure}

One can see that scaling law works for the Stokes - Einstein
coefficient too \cite{footnote}. The apparent choice for the
particle size $\sigma$ is the Barker diameter. For soft sphere
system the Barker diameter is defined as \cite{hansmcd,barhend}:

\begin{equation}
 d_B=\sigma \left(\frac{\varepsilon}{k_BT}\right)^{1/n}
 \Gamma\left(\frac{n-1}{n}\right),
\end{equation}
i.e. for our case $c \sim T^{-1/12}$. From this one can conclude
that the Barker diameter should be used as a characteristic
particle size. Fig. 10 shows $c/T^{1/12}$. One can see that the
Stokes - Einstein coefficient becomes constant and equals $2$ with
this normalization.

\begin{figure}
\includegraphics[width=7cm]{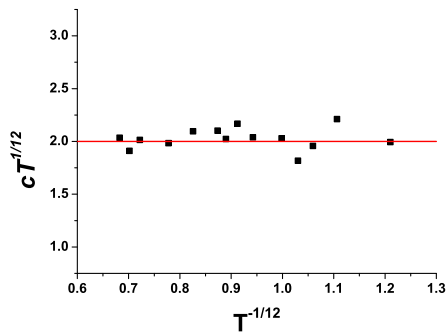}%
\caption{\label{fig:fig8} Stokes - Einstein coefficient
$c/T^{1/12}$.}
\end{figure}

\subsection{Isoviscosity Lines}

As it was mentioned in the introduction, the shape of the
isoviscosity lines is still unclear. There are two suggestions
claiming that the viscosity is constant along the isochore or
along the melting curve, but both of them contradict to the
experimental data \cite{brlyap}. In this respect it is interesting
to see the behavior of isoviscosity lines in the soft sphere
system as a generic model for more complex liquids.

Figs. 11 and 12 present the isoviscosity lines in comparison with
the viscosity along the melting line and along the isochores. One
can see that the slope of the isoviscosity lines is higher then
the viscosity along the melting line, but lower then the one along
the isochores. One of us discussed the behavior of the
isoviscosity lines in the previous publication \cite{brlyap}.
Based on the experimental data for mercury and argon, it was
suggested that the isoviscosity lines are located between the
melting line and isochore. From this one can see that the
simulation results qualitatively correspond to the experimental
data.

\begin{figure}
\includegraphics[width=9cm]{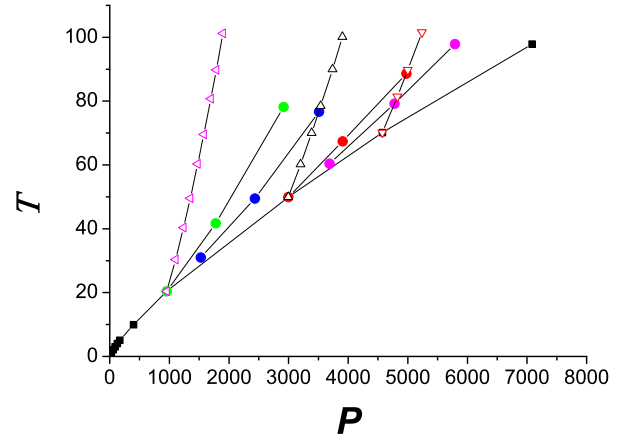}%

\caption{\label{fig:fig9b} Isoviscosity lines compared with the
viscosity along the melting line and isochores. Squares - melting
line, opened triangles - isochores (from left to right:
$\rho=2.43, 3.06$ and $3.33$), filled circles - isoviscosity lines
(from left to right: $\eta=41.61, 55.32, 77.94$ and $88.86$).
(Color online)}
\end{figure}

\begin{figure}
\includegraphics[width=9cm]{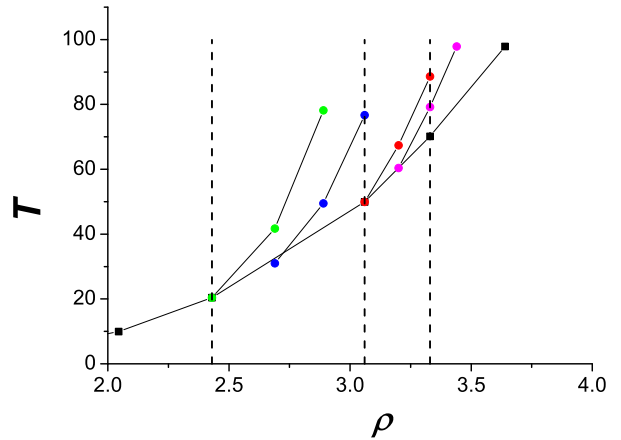}%

\caption{\label{fig:fig9b} Isoviscosity lines compared with the
viscosity along the melting line and isochores. Squares - melting
line, filled circles - isoviscosity lines (from left to right:
$\eta=41.61, 55.32, 77.94$ and $88.86$). Note that the isochores
are vertical lines in these coordinates (dashed lines here).
(Color online)}
\end{figure}

\subsection{Excess Entropy Scaling}

It is important from both theoretical and practical points of view
to establish some relationships between dynamic and thermodynamic
properties of liquids. One of the most common relations was
developed by Rosenfeld \cite{rosenfeld,ros1}. According to
Rosenfeld, one can define some reduced transport coefficients
which exponentially depend on the excess entropy of the liquid.
The reduced diffusion $D^*$ and viscosity $\eta ^*$ have the
following forms:

\begin{equation}
 D^*=D \frac{\rho ^{1/3}}{(k_BT/m)^{1/2}}
\end{equation}
\begin{equation}
 \eta^*=\eta \frac{\rho^{-2/3}}{(mk_BT)^{1/2}},
\end{equation}
and their dependence on the excess entropy
$S_{ex}=(S-S_{id})/{(Nk_B)}$ looks like
\begin{equation}
  X=a_X \cdot e^{b_X S_{ex}},
\end{equation}
where $X$ is the transport coefficient ($D$ or $\eta$ in the
present case), and $a_X$ and $b_X$ are the constants which depend
on the studied property \cite{rosenfeld,ros1}.

Very often one can replace the excess entropy by a pair
contribution to it, which can be computed via the relation
\begin{equation}
 s_{2}=-\frac{1}{2} \rho \int
 d\textbf{r}[g(\textbf{r})\ln(g(\textbf{r}))-(g(\textbf{r})-1)],
\end{equation}
where $g(\textbf{r})$ is radial distribution function of the
liquid.

In the present work we verify the scaling relations for the
diffusion coefficient and viscosity. To do this, we calculate the
pair and full excess entropies along the isochore $\rho=1.51$. The
pair excess entropy is calculated from the radial distribution
function. For computing the full excess entropy, the following
procedure is applied. As we mentioned above, the free energies of
the system at $T=20.0$ were calculated for determining the melting
point. After that we computed the energy along the isochore
$\rho=1.51$. Having free energy at $T_0=20.0$ and the temperature
dependence of internal energy, one can find the free energy along
this isochore:
$\frac{F(T)}{k_BT}-\frac{F_{0}}{k_BT_{0}}=\int_{T_0}^{T}
{d(\frac{1}{T'})U(T')}$.  The excess entropy along the isochore is
obtained by differentiating the free energy with respect to
temperature.

\begin{figure}
\includegraphics[width=9cm]{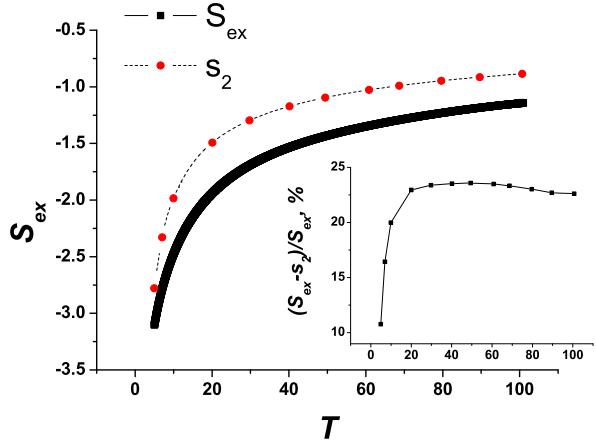}%
\caption{\label{fig:fig8} Full (squares) and pair (circles) excess
entropy of the soft spheres along the isochore $\rho=1.51$. The
inset represents the relative error $\frac{S_{ex}-s_2}{S_{ex}}$ in
percent. (Color online)}
\end{figure}

Fig.13 presents the excess entropy and pair contribution to it
along the isochore $\rho=1.51$. One can see that at some
temperature the relative difference of the entropies becomes
approximately constant. In the case of the exponential dependence,
it gives a constant shift of the curves.

\begin{figure}
\includegraphics[width=9cm]{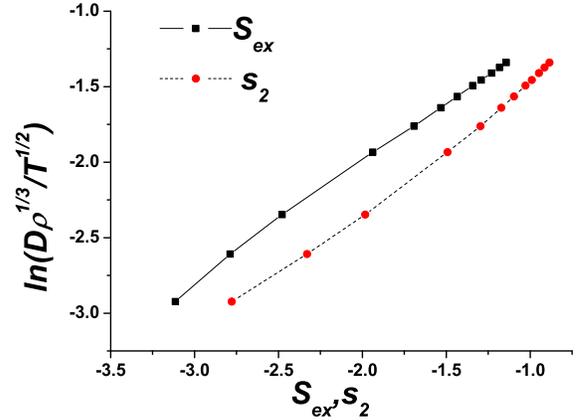}%
\caption{\label{fig:fig8} Logarithm of reduced diffusion $D^*$ as
a function of full (squares) and pair (circles) excess entropy.
(Color online)}
\end{figure}

\begin{figure}
\includegraphics[width=9cm]{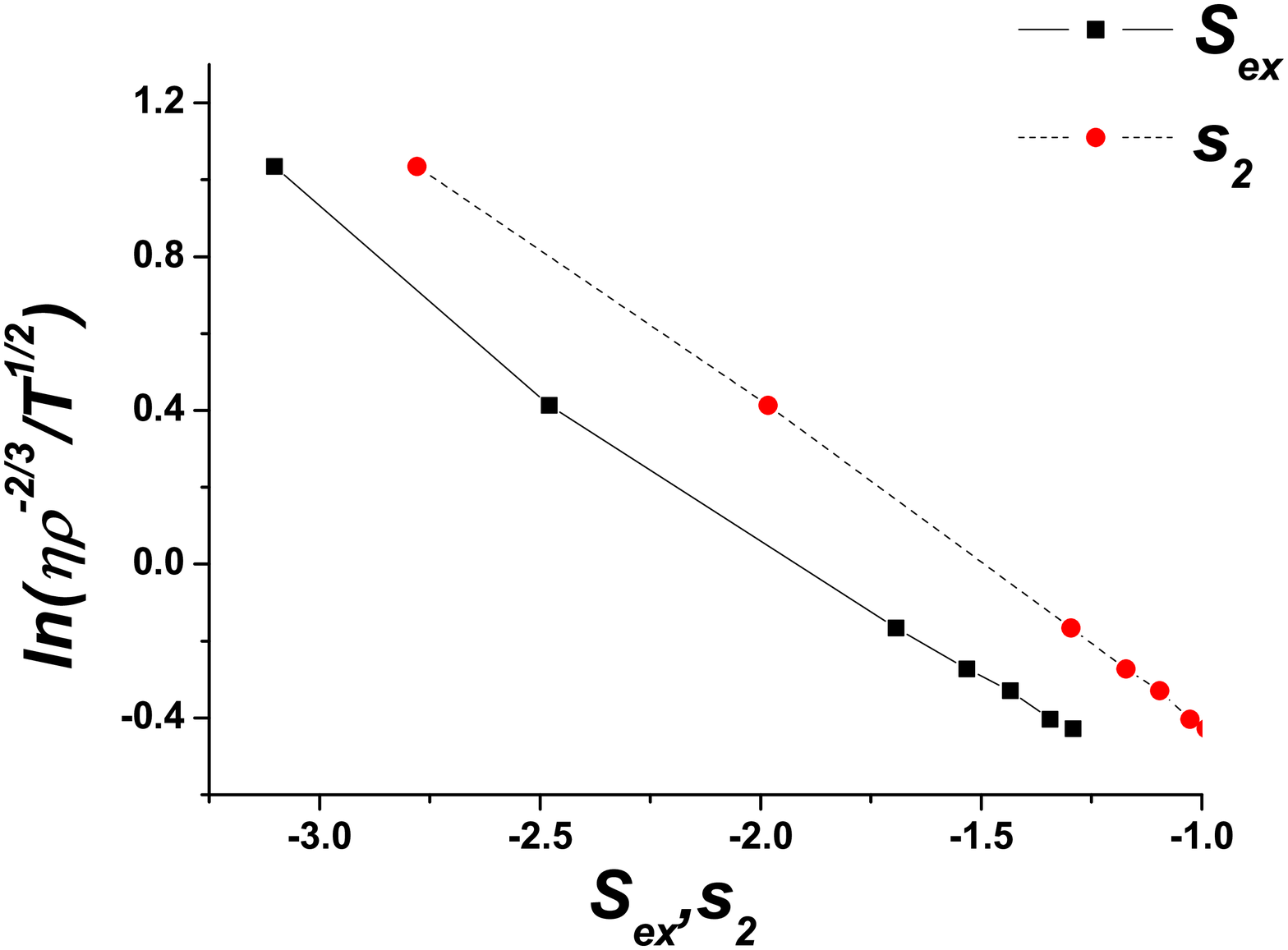}%
\caption{\label{fig:fig8} Logarithm of reduced viscosity $\eta^*$
as a function of full (squares) and pair (circles) excess entropy.
(Color online)}
\end{figure}

Figs. 14 and 15 show the logarithm of the Rosenfeld scaled
diffusion and viscosity along this isochore. As can be seen from
these plots, the scaling relation is fulfilled with good accuracy
in a wide range of temperatures. The curves corresponding to the
full excess entropy and to the pair contribution to it have the
same shape but shifted with respect to each other, which
corresponds to the intuitive conclusion described above.

In conclusion to this subsection, we would like to make some
remarks. First of all, recently another mechanism for the origin
of the excess entropy scaling has been suggested \cite{dyre}.
According to this work, the scaling law is valid for the liquids
with strong energy-virial correlations (the authors call them
"strongly correlated liquids"). The class of strongly correlated
liquids includes many different systems, like, for example, a
Lennard-Jones liquid, Kob-Andersen mixture, square-well liquid and
many others. The soft spheres considered in this article are
exactly the strong correlated liquids \cite{dyre2}.

On the other hand, the applicability of the entropy scaling to
not-strongly-correlated liquids is questionable. It seems that in
the case of the existence of thermodynamic anomalies in the system
\cite{FFGRS2008,GFFR2009}, Rosenfeld scaling does not work. In
Ref. \onlinecite{FGR2010} it was shown that for Herzian spheres,
Gauss Core Model, and soft repulsive shoulder potential, the
Rosenfeld entropy scaling breaks down. These systems demonstrate
the diffusion anomalies at low temperatures: the diffusion
increases with increasing density or pressure. It is shown that
for the first two systems, which belong to the class of bounded
potentials, the Rosenfeld scaling formula is valid only in the
infinite temperature limit where there are no anomalies. For the
soft repulsive shoulder, the scaling formula is valid already at
sufficiently low temperatures, however, out of the anomaly range.

\subsection{Influence of Attraction}

Finally we would like to study the influence of attractive forces
on the behavior of the viscosity. To do this, we simulate the
Lennard-Jones like system. The system has the potential function:

\begin{equation}
  \Phi_{LJ}=\varepsilon \cdot [(\frac{\sigma}{r})^{12}-(\frac{\sigma}{r})^6]
\end{equation}
with the parameters $\varepsilon=1$ and $\sigma=1$.

We measure the viscosity of both LJ systems at the isochore
$\rho=1.51$. The resulting curves are shown in Figs. 16a and 16b.
One can see from these figures that the difference takes place
only in the low temperature region and should be concerned to the
\textit{pp} contribution while at high temperatures where
\textit{kk} contribution is dominant the viscosity of soft spheres
and LJ system have very close magnitudes.

\begin{figure}
\includegraphics[width=7cm]{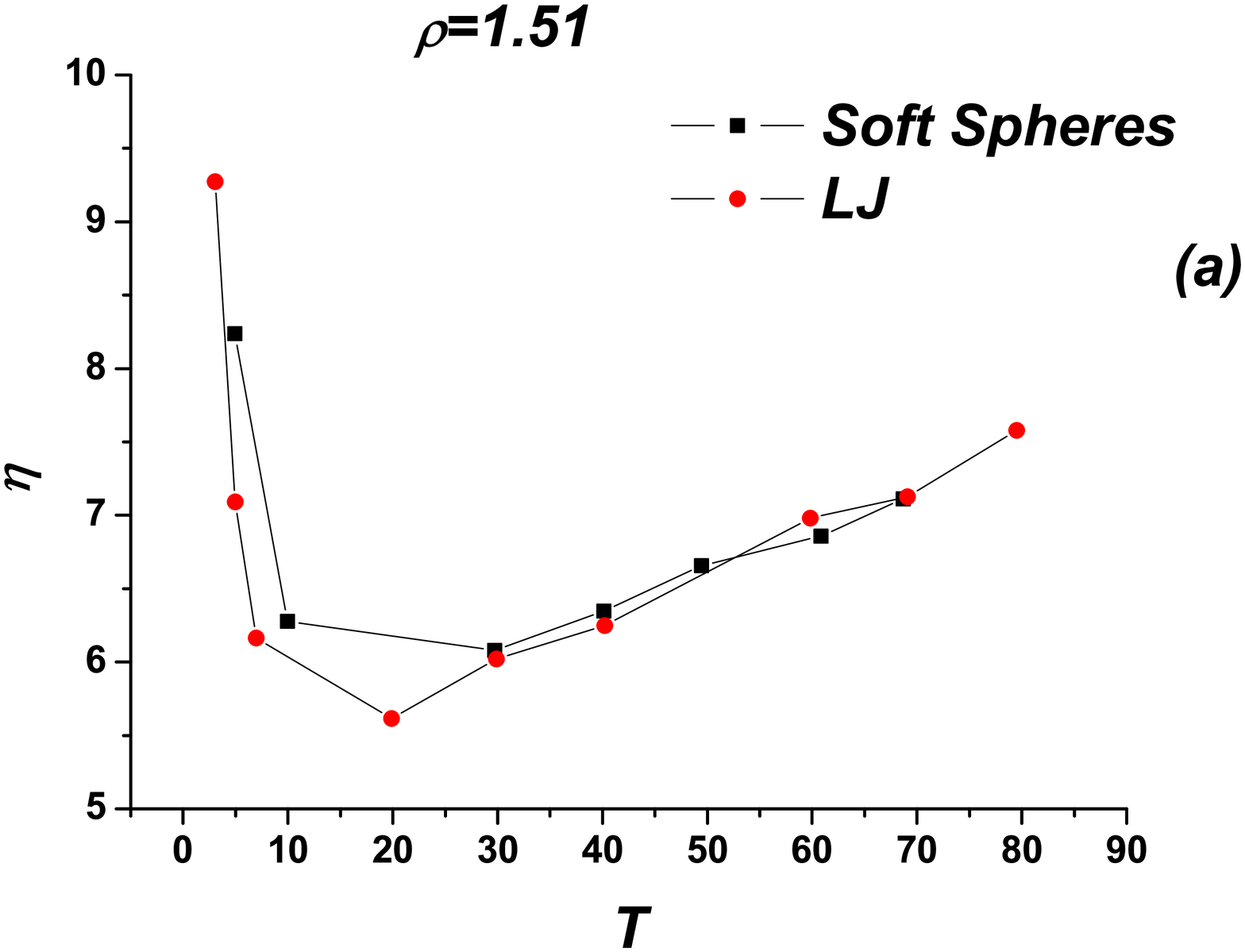}%

\includegraphics[width=7cm]{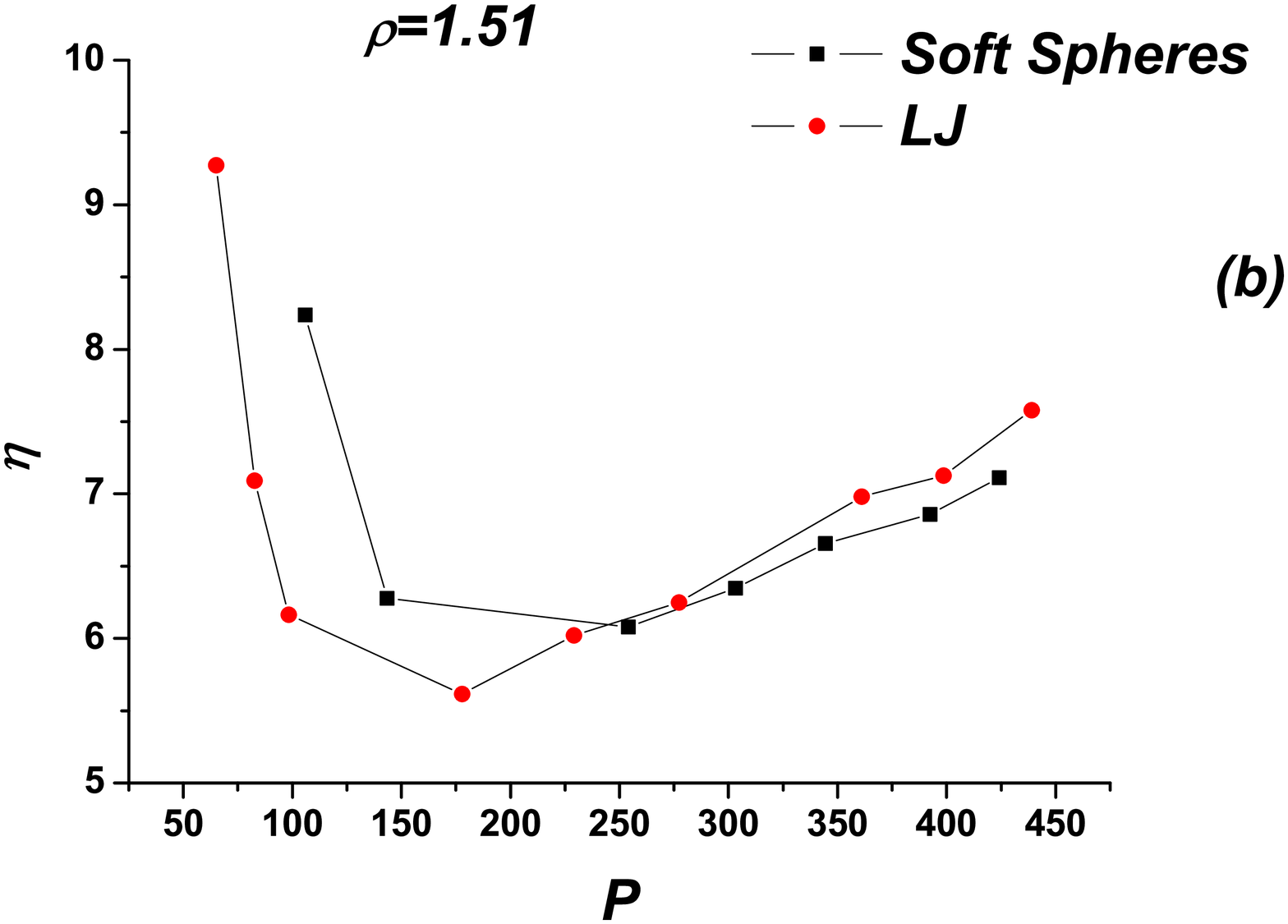}%
\caption{\label{fig:fig10b} Viscosity of the soft sphere and
Lenard-Jones systems along the isochore $\rho=1.51$. Squares -
soft spheres, and circles - LJ. (Color online)}
\end{figure}

Based on these results, one can expect that in the case of real
liquids the viscosity can demonstrate complex behavior on
compression, since they are characterized by complex non pair
potentials which can even be state dependent. The results for the
soft sphere system can directly be applied only for the rare gas
liquids, although even in this case the attraction is important,
as is seen from Figs. 16a and b.

As one can see from the figures the difference between the soft
spheres and Lennard-Jones system takes place only in the low
temperature regime. One can attribute it to the difference in
potential - potential contribution to the viscosity while at high
temperature the influence of the kinetic - kinetic term is not so
strongly affected by the interatomic potential.

\section{IV. Conclusions}

In conclusion we have carried out a comprehensive study of the
viscosity and diffusivity of the soft sphere system. We showed the
validity of the scaling laws along the melting line. We proposed
the scaling law for the Stokes - Einstein coefficient and verified
it. The influence of different contributions to the viscosity was
discussed. It was shown that the viscosity is strongly growing
along the melting line, however, this growth does not stimulate
the glass transition because the relaxation time is decreasing. We
calculated the shape of the isoviscosity lines and compared them
with the viscosity along the melting line and isochores. The
results we obtained are qualitatively similar to the experimental
findings. The validity of the Rosenfeld excess entropy scaling was
also verified. Interestingly, although the qualitative behavior of
viscosity and diffusivity is rather different, the excess entropy
scaling describes both coefficients with high accuracy. Finally,
the influence of the attractive forces was verified via the
comparison to the Lenard-Jones system. It was shown that the
attraction is important only at sufficiently low temperatures.

\bigskip

\begin{acknowledgments}
We thank S. M. Stishov, E. E. Tareyeva and A. G. Lyapin for
stimulating discussions. Y.F. thanks the Joint Supercomputing
Center of the Russian Academy of Sciences for computational power.
The work was supported in part by the Russian Foundation for Basic
Research (Grant No 08-02-00781).
\end{acknowledgments}


\begin{thebibliography}{99}
\bibitem{footnote} The noise in the Stokes - Einstein coefficient is because of
the errors in the viscosity computation.
\end{thebibliography}

\end{document}